\documentclass[9pt]{IEEEtran}

\IEEEoverridecommandlockouts
\usepackage{cite}
\usepackage{amsmath,amssymb,amsfonts}
\usepackage{algorithmic}
\usepackage{graphicx}
\usepackage{textcomp}
\usepackage{xcolor}
\usepackage{url}
\usepackage{hyperref} 
\usepackage{cite}
\usepackage{booktabs}
\def\BibTeX{{\rm B\kern-.05em{\sc i\kern-.025em b}\kern-.08em
    T\kern-.1667em\lower.7ex\hbox{E}\kern-.125emX}}

\usepackage{caption} 
\usepackage{tabularx}
\usepackage{float}
\usepackage{array}  
\usepackage{booktabs} 
\usepackage{multirow}
\usepackage{makecell}
\usepackage{arydshln} 
\usepackage{color, url}
\usepackage{bbding}
\usepackage{hyperref}
\usepackage{amsfonts,amssymb} 
\usepackage{subfigure} 
\usepackage{bbding}  
 
\usepackage{graphicx}

\title{\includegraphics[height=1.8em]{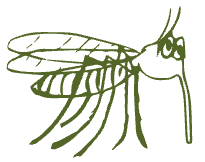}\hspace{0.8em}BioDCASE 2026 Challenge Baseline for Cross-Domain Mosquito Species Classification}

\author{
    \IEEEauthorblockN{
        Yuanbo Hou$^1$, 
        Vanja Zdravkovic$^1$,
        Marianne Sinka$^2$,
        Yunpeng Li$^3$,  
        Wenwu Wang$^4$, 
        Mark D. Plumbley$^3$,\\
        Kathy Willis$^2$,
        Stephen Roberts$^1$
    }

    $^1$Machine Learning Research Group, University of Oxford; \\
    $^2$Oxford Long-Term Ecology Laboratory, University of Oxford; \\ 
    $^3$King’s College London; 
    $^4$Centre for Vision, Speech and Signal Processing, University of Surrey 
}

\begin{document}

\maketitle

\begin{abstract}
Mosquito-borne diseases affect more than one billion people each year and cause close to one million deaths. Traditional surveillance methods rely on traps and manual identification that are slow, labor-intensive, and difficult to scale. Audio-based mosquito monitoring offers a non-destructive, lower-cost, and more scalable complement to trap-based surveillance, but reliable species classification remains difficult under real-world recording conditions. Mosquito flight tones are narrow-band, often low in signal-to-noise ratio, and easily masked by background noise, and recordings for several epidemiologically relevant species remain limited, creating pronounced class imbalance. Variation across devices, environments, and collection protocols further increases the difficulty of robust classification. Such variation can cause models to rely on domain-specific recording artefacts rather than species-relevant acoustic cues, which makes transfer to new acquisition settings difficult. The BioDCASE 2026 Cross-Domain Mosquito Species Classification (CD-MSC) challenge is designed around this deployment problem by evaluating performance on both seen and unseen domains. This paper presents the official baseline system and evaluation pipeline as a simple, fully reproducible reference for the CD-MSC challenge task. The baseline uses log-mel features and a multitemporal resolution convolutional neural network (MTRCNN) with species and auxiliary domain outputs, together with complete training and test scripts. Official evaluation ranking is based on balanced accuracy (BA) on unseen domains, while balanced accuracy on seen domains and the gap between them are reported to measure robustness under domain shift. The baseline system performs strongly on seen domains but degrades markedly on unseen domains, showing that cross-domain generalisation, rather than within-domain recognition, is the central challenge for practical mosquito species classification from multi-source bioacoustic recordings.
\end{abstract}


\section{Introduction}
 
Mosquito-borne diseases remain a major global health challenge. Traditional mosquito surveillance relies on traps and manual species identification, which are slow, labour-intensive, difficult to scale, and can expose field workers to infection risk \cite{who_vector_borne}. Audio-based mosquito monitoring offers a non-destructive, lower-cost, and more scalable complement to this workflow. Flight sounds can be captured with low-cost sensors, including mobile phones, supporting broader monitoring in settings where conventional surveillance is harder to deploy \cite{sinka2021humbug,kiskin2humbugdb,mukundarajan2017using,paim2024acoustic}. Practical deployment potential is therefore clear, but reliable mosquito species classification from audio remains difficult under real recording conditions.

Cross-domain robustness is the central bottleneck. Mosquito flight sounds are narrow-band, often weak, and frequently recorded at low signal-to-noise ratio, which makes them vulnerable to masking by background noise \cite{kiskin2humbugdb,drbiol_icassp2026}. Recordings for several epidemiologically relevant species also remain limited, creating pronounced class imbalance \cite{kiskin2humbugdb}. Performance is further affected by variation across devices, environments, and collection protocols, while within-species acoustic variation can arise from sex, size, age, and temperature \cite{kiskin2humbugdb,mukundarajan2017using,paim2024acoustic,loh2024differences,drbiol_icassp2026}. Under these conditions, a model can appear effective in familiar domains while relying on recording-specific artefacts rather than stable species-relevant acoustic cues, which leads to poor transfer to new acquisition settings \cite{drbiol_icassp2026}. Therefore, practical mosquito species classification requires not only accurate recognition within known conditions but also reliable generalisation across domains.

The BioDCASE 2026 Cross-Domain Mosquito Species Classification (CD-MSC) challenge\footnote{BioDCASE 2026 CD-MSC challenge homepage: \textcolor{blue}{\url{https://biodcase.github.io/challenge2026/task5}}} is designed around this deployment problem. The final evaluation setting includes both seen and unseen domains that differ in geography, device type, or acoustic environment. Official ranking is based on balanced accuracy on unseen domains, while balanced accuracy (BA) on seen domains and the gap between them are reported to characterise robustness under domain shift. CD-MSC is framed around three linked questions: how strongly performance degrades in new domains, whether models retain species-relevant cues or rely on domain-specific cues, and whether seen--unseen diagnostic gaps can distinguish systems that generalise well from those that deteriorate sharply. The challenge tests not only whether a system can recognise mosquito species, but also whether learned representations remain species-relevant when acquisition conditions change. This paper presents the official baseline system and evaluation pipeline as a simple, fully reproducible reference for the CD-MSC task, establishing a clear benchmark for future domain-robust mosquito species classification methods.

\section{Challenge Setting and Baseline Development Split}

This section defines the official cross-domain evaluation objective of CD-MSC, describes the released development data and split statistics used by the baseline, and clarifies how the released baseline system should be interpreted relative to the final challenge setting.

\subsection{Task Objective and Evaluation Metrics}

A domain in CD-MSC denotes an acquisition condition associated with the recording source, including differences in location, device, or acoustic environment. The official task targets mosquito species classification under domain shift rather than pooled recognition under matched conditions. In the final challenge evaluation, each species is assessed using recordings drawn from both seen and unseen domains, so that performance can be reported separately for source-like and shifted conditions. The official evaluation focuses on cross-domain generalisation rather than within-domain recognition alone.

The official evaluation reports balanced accuracy on seen domains, denoted $\mathrm{BA}_{\mathrm{seen}}$, and balanced accuracy on unseen domains, denoted $\mathrm{BA}_{\mathrm{unseen}}$. 
For class $c \in \{1,\ldots,C\}$, let $N^{\text{unseen}}_c$ be the number of clips from unseen domains with true label $c$, and let $\mathrm{TP}^{\text{unseen}}_c$ be the number correctly classified. 
The metric $\mathrm{BA}_{\mathrm{unseen}}$ is cacluated as
\[ 
\mathrm{BA}_{\text{unseen}} = \frac{1}{C} \sum_{c=1}^{C} \frac{\mathrm{TP}^{\text{unseen}}_c}{N^{\text{unseen}}_c}.
\]
Let $N^{\text{seen}}_c$ and $\mathrm{TP}^{\text{seen}}_c$ denote the corresponding quantities for clips from domains in the training set,
$
\mathrm{BA}_{\text{seen}} = \frac{1}{C} \sum_{c=1}^{C} \frac{\mathrm{TP}^{\text{seen}}_c}{N^{\text{seen}}_c}$. 

$\mathrm{BA}_{\mathrm{unseen}}$ is the primary indicator of cross-domain generalisation. The challenge reports the {Domain Shift Gap} (DSG) between them,
\begin{equation}
\mathrm{DSG} = |\mathrm{BA}_{\mathrm{unseen}} - \mathrm{BA}_{\mathrm{seen}}|
\end{equation}
which characterises the extent to which performance changes under domain shift. Under this setting, a strong system should not only classify mosquito species accurately, but should also retain species-relevant behaviour when acquisition conditions change, rather than relying mainly on source-domain artefacts.

 \subsection{Released Development Dataset and Split Statistics}

The baseline is designed as a simple and reproducible reference implementation for this task. The released development dataset contains 271{,}380 clips in total, corresponding to 218{,}388.40 seconds (60.66 hours), as shown in Tables~\ref{tab:dev_count_by_domain_species} and~\ref{tab:dev_duration_by_domain_species}. Each audio file follows the naming format \texttt{S\_<speciesID>\_D\_<domainID>\_<clipIndex>}, so both species identity and domain identity are directly accessible from the filename. Participants can therefore recover domain membership without additional metadata lookup and construct alternative domain-aware development splits if needed. The dataset includes nine target species, \textit{Ae. aegypti}, \textit{Ae. albopictus}, \textit{Cx. quinquefasciatus}, \textit{An. gambiae}, \textit{An. arabiensis}, \textit{An. dirus}, \textit{Cx. pipiens}, \textit{An. minimus}, and \textit{An. stephensi}, from five domains.

For the CD-MSC baseline, the development dataset is divided into a trainval set of 244{,}163 clips and a held-out test set of 27{,}217 clips. The held-out test set in Table~\ref{tab:test_distribution}, contains both seen-domain and unseen-domain recordings, enabling the calculation of $\mathrm{BA}_{\mathrm{seen}}$, $\mathrm{BA}_{\mathrm{unseen}}$, and $\mathrm{DSG}$. A validation set is then derived from the trainval set by random species-stratified sampling, using a validation fraction of 12.498\%. This yields 213{,}647 training clips and 30{,}516 validation clips. The species-domain composition of the trainval set is summarised in Table~\ref{tab:trainval_distribution}.

The trainval set and held-out test set remain highly uneven across species-domain combinations. Such asymmetry means that pooled clip-level accuracy can conceal substantial differences between source-like and shifted conditions, which further motivates the use of balanced accuracy and separate reporting on seen and unseen domains in the official evaluation.

\begin{table}[t]
\centering
\caption{Number of clips in the full development dataset. Rows correspond to domains and columns correspond to species.}
\label{tab:dev_count_by_domain_species}
\scriptsize
\setlength{\tabcolsep}{3pt}      
\renewcommand{\arraystretch}{1.05} 
\resizebox{\columnwidth}{!}{%
\begin{tabular}{lrrrrrrrrrr}
\hline
Dataset & Ae.aeg & Ae.alb & Cx.qui & An.gam & An.ara & An.dir & Cx.pip & An.min & An.ste & Total \\
\hline
D1 & 123 & 79 & 672 & 818 & 1820 & 87 & 66 & 200 & 200 & 4065 \\
D2 & 0 & 419 & 0 & 0 & 0 & 0 & 66 & 99 & 200 & 784 \\
D3 & 192 & 19 & 0 & 0 & 0 & 0 & 68 & 200 & 200 & 679 \\
D4 & 22 & 0 & 13 & 0 & 0 & 40 & 0 & 51 & 74 & 200 \\
D5 & 81250 & 18000 & 71371 & 46180 & 19297 & 0 & 29554 & 0 & 0 & 265652 \\
\hline
Total & 81587 & 18517 & 72056 & 46998 & 21117 & 127 & 29754 & 550 & 674 & 271380 \\ 
\hline
\end{tabular}%
}
\end{table}

\begin{table}[b]
\centering
\caption{Total duration in seconds in the full development dataset. Rows correspond to domains and columns correspond to species.}
\label{tab:dev_duration_by_domain_species}
\scriptsize
\setlength{\tabcolsep}{3pt}      
\renewcommand{\arraystretch}{1.05} 
\resizebox{\columnwidth}{!}{%
\begin{tabular}{lrrrrrrrrrr}
\hline
Dataset & Ae.aeg & Ae.alb & Cx.qui & An.gam & An.ara & An.dir & Cx.pip & An.min & An.ste & Total \\
\hline
D1 & 1332.65 & 42.32 & 7418.64 & 2978.16 & 18322.92 & 748.09 & 132.66 & 424.85 & 381.63 & 31781.92 \\
D2 & 0.00 & 1280.58 & 0.00 & 0.00 & 0.00 & 0.00 & 132.66 & 198.99 & 402.00 & 2014.23 \\
D3 & 5909.57 & 580.46 & 0.00 & 0.00 & 0.00 & 0.00 & 136.68 & 402.00 & 402.00 & 7430.71 \\
D4 & 1737.07 & 0.00 & 758.19 & 0.00 & 0.00 & 1775.24 & 0.00 & 2944.18 & 2586.10 & 9800.78 \\
D5 & 51187.50 & 11340.00 & 44963.73 & 29093.40 & 12157.11 & 0.00 & 18619.02 & 0.00 & 0.00 & 167360.76 \\
\hline
Total & 60166.79 & 13243.36 & 53140.56 & 32071.56 & 30480.03 & 2523.33 & 19021.02 & 3970.02 & 3771.73 & 218388.40 \\
\hline 
\end{tabular}%
}
\end{table}

\begin{table}[t]
\centering
\caption{Species-domain distribution of the trainval pool set.}
\label{tab:trainval_distribution}
\scriptsize
\setlength{\tabcolsep}{2.5pt}      
\renewcommand{\arraystretch}{1.05} 
\resizebox{\columnwidth}{!}{%
\begin{tabular}{lrrrrrrrrrr}
\hline
Dataset & Ae.aeg & Ae.alb & Cx.qui & An.gam & An.ara & An.dir & Cx.pip & An.min & An.ste & Total \\
\hline
D1 & 111 & 73 & 0 & 0 & 0 & 87 & 59 & 200 & 200 & 730 \\
D2 & 0 & 0 & 0 & 0 & 0 & 0 & 60 & 0 & 200 & 260 \\
D3 & 0 & 17 & 0 & 0 & 0 & 0 & 0 & 200 & 200 & 417 \\
D4 & 20 & 0 & 12 & 0 & 0 & 0 & 0 & 51 & 0 & 83 \\
D5 & 73297 & 16575 & 64838 & 42298 & 19005 & 0 & 26660 & 0 & 0 & 242673 \\
\hline
\end{tabular}%
}
\end{table}

\begin{table}[b]
\centering
\caption{Species-domain distribution of the test set.}
\label{tab:test_distribution}
\scriptsize
\setlength{\tabcolsep}{2.5pt}      
\renewcommand{\arraystretch}{1.05} 
\resizebox{\columnwidth}{!}{%
\begin{tabular}{lrrrrrrrrrr}
\hline
Dataset & Ae.aeg & Ae.alb & Cx.qui & An.gam & An.ara & An.dir & Cx.pip & An.min & An.ste & Total \\
\hline
D1 & 12 & 6 & 672 & 818 & 1820 & 0 & 7 & 0 & 0 & 3335 \\
D2 & 0 & 419 & 0 & 0 & 0 & 0 & 6 & 99 & 0 & 524 \\
D3 & 192 & 2 & 0 & 0 & 0 & 0 & 68 & 0 & 0 & 262 \\
D4 & 2 & 0 & 1 & 0 & 0 & 40 & 0 & 0 & 74 & 117 \\
D5 & 7953 & 1425 & 6533 & 3882 & 292 & 0 & 2894 & 0 & 0 & 22979 \\
\hline
\end{tabular}%
}
\end{table}

\section{Baseline System} 

The baseline uses log-mel spectrograms and a lightweight tiny convolutional classifier to provide a simple and reproducible benchmark for CD-MSC. Input audio is resampled to 8 kHz and converted to log-mel spectrograms. Feature extraction uses a 512-point FFT with a hop size of 10 ms, producing 64 mel bands over the 0--4 kHz range with a frequency resolution of 15.625 Hz. This setting provides a practical balance between representing the low-frequency structure of mosquito flight sounds and keeping computation efficient. Waveforms are peak-normalised before feature extraction to stabilise input scaling, and feature standardisation uses the mean and standard deviation estimated from the training set only, avoiding information leakage from validation or test data.
 
The classifier is a lightweight multitemporal resolution convolutional neural network (MTRCNN) \cite{hou2025sound}, which consists
of 3 parallel branches with convolutional kernel sizes of (3, 3), (5, 5), and (7, 7), and each branch includes three convolutional layers with a hybrid dilated convolution scheme. MTRCNN can handle audio clips with varying lengths with a minimum length of 1.10s \cite{hou2025sound}. Following the baseline design, the model contains a primary species-classification head and an auxiliary domain-classification head. Species prediction remains the main task objective. The auxiliary domain head provides additional supervised training signals from the available domain labels, but it is not used to enforce domain invariance, adversarial alignment, or explicit domain adaptation. Variable-length clips are handled through dynamic padding and length-aware masking, which allows the same architecture to process clips of different durations. The model contains 0.22 million trainable parameters \cite{hou2025sound}, keeping the baseline lightweight and straightforward as a transparent reference implementation for participants.
Source code and models are available on the homepage (\textcolor{blue}{\url{https://github.com/Yuanbo2020/CD-MSC}}).

\section{Baseline Experiments and Results}

\subsection{Experimental setup}

The baseline is trained under a fixed and reproducible setup. Optimisation uses AdamW \cite{adamw} with a learning rate of 0.001. The training batch size is 64, and evaluation uses a batch size of 8. Training runs for at most 100 epochs. Early stopping is activated after epoch 10 with a patience of 5 epochs. Model selection is based on validation species balanced accuracy.

To regularise clip length during training, recordings longer than 2.0\,s are randomly cropped, corresponding to 200 frames under the feature configuration. Validation, test, and prediction use the full clip. The released code supports repeated experiments with fixed random seeds. The default setup uses ten seeds: 42, 3407, 1234, 2023, 2024, 1024, 2048, 4096, 8192, and 10086. Reported results are given as mean $\pm$ standard deviation across these runs. 
 
For each run, the baseline stores both the best-validation checkpoint and the final checkpoint. The best-validation checkpoint is selected using validation species balanced accuracy and is treated as the primary reference model in the following results. Final-checkpoint results are retained only as a secondary reference to show whether the observed cross-domain pattern is stable.

\subsection{Cross-Domain MSC results and analysis} 

Table~\ref{tab:official_results} reports the cross-domain results of the released baseline. The best-validation checkpoint achieves $\mathrm{BA}_{\mathrm{seen}} = 0.8806$ and $\mathrm{BA}_{\mathrm{unseen}} = 0.1751$, the corresponding DSG is $0.7055$. The released baseline performs strongly on seen domains but degrades sharply on unseen domains. Therefore, the central difficulty of CD-MSC is not recognition under source-like conditions, but transfer to new recording conditions.

\begin{table}[b]
\centering 
\caption{Performance of the baseline checkpoints on seen-domain and unseen-domain samples from the test set of the development dataset, reported over 10 runs.}
\label{tab:official_results}
\small
\setlength{\tabcolsep}{4pt}       
\renewcommand{\arraystretch}{1.05} 
\resizebox{\columnwidth}{!}{%
\begin{tabular}{lccc}
\hline
Checkpoint & $\mathrm{BA}_{\mathrm{seen}}$ & $\mathrm{BA}_{\mathrm{unseen}}$ & DSG \\
\hline
Best-validation & $0.8806 \pm 0.0108$ & $0.1751 \pm 0.0197$ & $0.7055 \pm 0.0248$ \\
Final & $0.8822 \pm 0.0097$ & $0.1704 \pm 0.0180$ & $0.7118 \pm 0.0235$ \\
\hline
\end{tabular}%
}
\end{table}

Table~\ref{tab:released_results} summarises the aggregate validation and test results of the released baseline on the development dataset. The final checkpoint gives slightly higher split-level species accuracy, whereas the best-validation checkpoint gives slightly higher split-level species balanced accuracy on the test set. The difference is small, indicating that checkpoint choice has only a limited effect on released-split aggregate performance. Therefore, selecting the model by validation species balanced accuracy remains a reasonable choice under the released development setup. All official cross-domain analyses below are reported for the best-validation checkpoint.

\begin{figure}[t]
\centering
\includegraphics[width=1\linewidth]{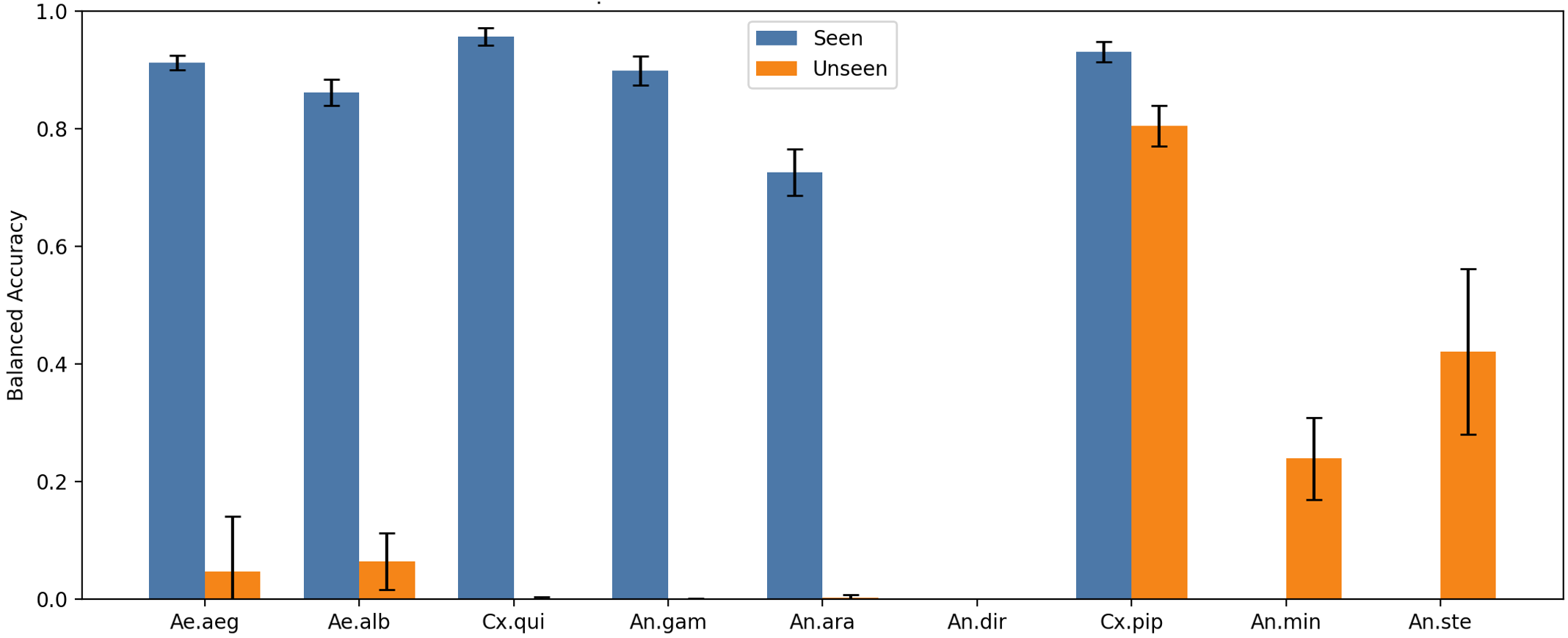} 
\caption{Per-species balanced accuracy of the best-validation checkpoint on the test set of the development dataset. For each species, $\mathrm{BA}_{\mathrm{seen}}$ and $\mathrm{BA}_{\mathrm{unseen}}$ are shown, with error bars indicating standard deviation over 10 runs.}
    \label{fig:species_seen_unseen}
\end{figure}

Fig.~\ref{fig:species_seen_unseen} shows the same per-species pattern. For most species, the gap between $\mathrm{BA}_{\mathrm{seen}}$ and $\mathrm{BA}_{\mathrm{unseen}}$ is large, indicating that cross-domain degradation is a broad property of the released baseline rather than an isolated effect in a few classes. \textit{Cx. pipiens} is the main exception, showing comparatively strong transfer across domains. This contrast further suggests that the severity of domain shift is species-dependent rather than uniform across the benchmark.

\begin{figure}[t]
\centering
\setlength{\abovecaptionskip}{0cm}   
	\setlength{\belowcaptionskip}{-0.1cm} 
\includegraphics[width=0.9\linewidth]{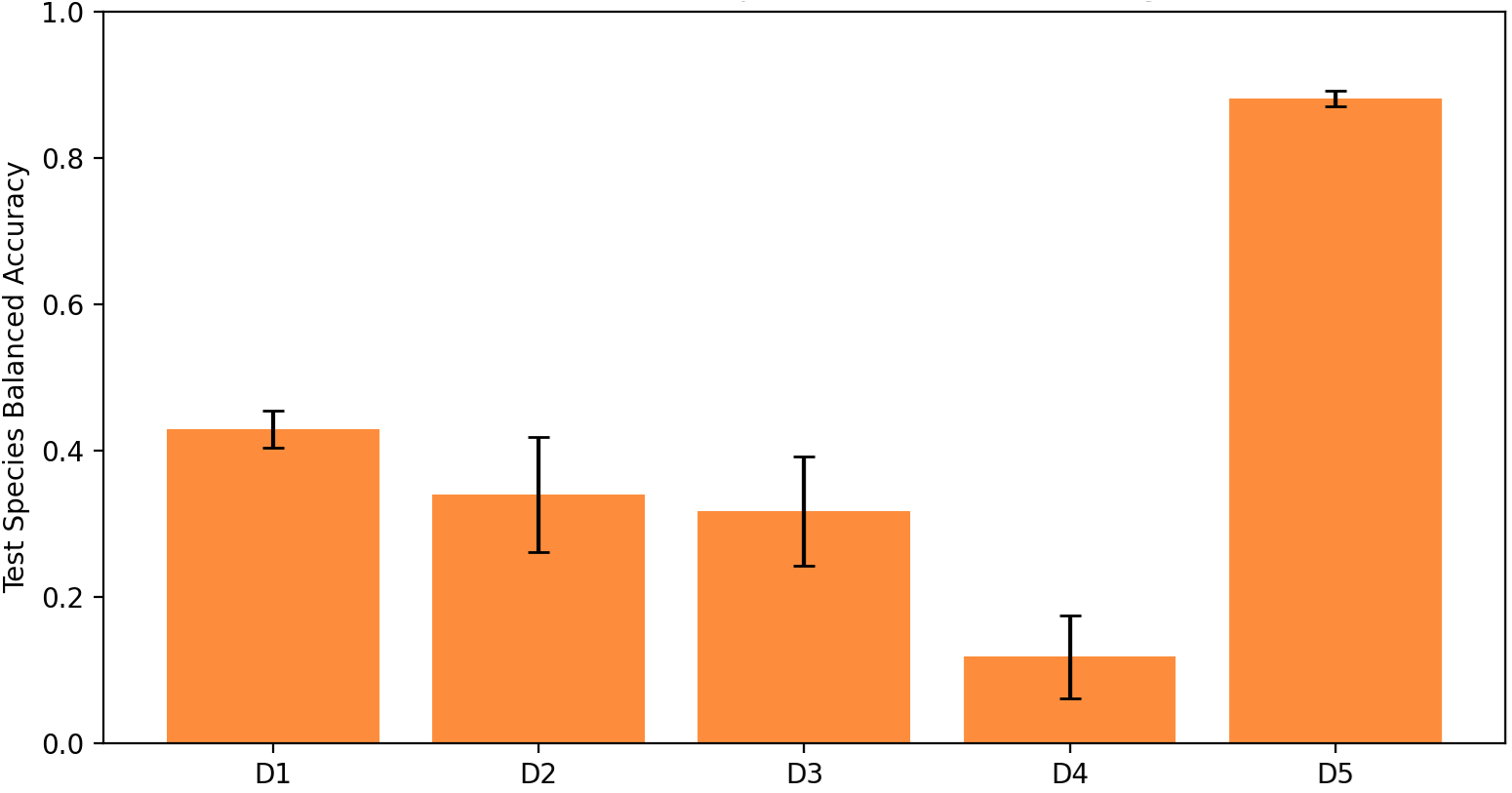}
\caption{Domain-level species balanced accuracy of the best-validation checkpoint on the test set of the development dataset, over 10 runs.}
\label{fig:domain_species}
\end{figure}

As shown in Fig.~\ref{fig:domain_species}, domain-level performance on the test set of the development dataset is highly uneven. D5 achieves the highest species balanced accuracy, whereas D4 is the most difficult domain. D1 to D3 lie between these extremes, with D1 performing better than D2 and D3. This pattern further highlights the difficulty of robust transfer across domains.

\begin{table}[b]
\centering 
\caption{Overall performance of the baseline checkpoints on the validation set and the test set of the development dataset, reported over 10 runs.}
\label{tab:released_results}
\footnotesize
\setlength{\tabcolsep}{2pt}        
\renewcommand{\arraystretch}{1.05} 
\resizebox{\columnwidth}{!}{%
\begin{tabular}{llcc}
\hline
Checkpoint & Metric & Validation & Test \\
\hline
\multirow{2}{*}{Best-validation}
& Species accuracy & $0.9056 \pm 0.0076$ & $0.7827 \pm 0.0068$ \\
& Species balanced accuracy & $0.8522 \pm 0.0108$ & $0.5411 \pm 0.0140$ \\
\hline
\multirow{2}{*}{Final}
& Species accuracy & $0.9096 \pm 0.0052$ & $0.7874 \pm 0.0034$ \\
& Species balanced accuracy & $0.8383 \pm 0.0176$ & $0.5380 \pm 0.0152$ \\
\hline
\end{tabular}%
}
\end{table}

\vspace{-2mm}
\section{Conclusion}

The BioDCASE 2026 Cross-Domain Mosquito Species Classification challenge establishes mosquito species recognition under domain shift as the central evaluation problem. Therefore, official evaluation prioritises unseen-domain balanced accuracy, rather than pooled performance under source-like conditions alone. The released baseline provides a simple, transparent, and reproducible reference implementation built on log-mel features, a lightweight MTRCNN classifier, and a fixed development setup. Results on the released setting show a clear and consistent pattern: the baseline performs strongly on seen-domain conditions but degrades sharply on unseen domains. Cross-domain transfer, rather than recognition under matched conditions, remains the main bottleneck of the task. All source code, models, and data are publicly available through the challenge homepage, supporting transparent comparison and future development on the CD-MSC benchmark.

\bibliographystyle{IEEEtran}
\bibliography{refs}

@inproceedings{kiskin2humbugdb,
  title={{HumBugDB}: A {Large-scale Acoustic Mosquito Dataset}},
  author={Kiskin, I. and Sinka, M. and Cobb, A. D. and others},
  booktitle={Proc. of NeurIPS}, 
  pages={58--68},
  year={2021}
}

@misc{who_vector_borne,
  title        = {Vector-borne diseases},
  author       = {{World Health Organization}},
  year         = {2024},
  howpublished = {\url{https://www.who.int/news-room/fact-sheets/detail/vector-borne-diseases}},
  note         = {Accessed: 2026-03-17}
}

@article{mukundarajan2017using,
  title={Using mobile phones as acoustic sensors for high-throughput mosquito surveillance},
  author={Mukundarajan, H. and Hol, F. J. and others},
  journal={Elife},
  volume={6},
  year={2017}
}

@article{paim2024acoustic,
  title={Acoustic identification of {Ae. aegypti} mosquitoes using smartphone apps and residual convolutional neural networks},
  author={Paim, K. and Rohweder, R.  and others},
  journal={Biomedical Signal Processing and Control},
  volume={95}, 
  year={2024}
}

@article{loh2024differences,
  title={Differences in male {Aedes aegypti and Aedes albopictus} hearing systems facilitate recognition of conspecific female flight tones},
  author={Loh, Y. and Xu, Y. and Lee, T. and Ohashi, T. and others},
  journal={Iscience},
  volume={27},
  number={7},
  year={2024}
}

@inproceedings{adamw,
title={{Decoupled Weight Decay Regularization}},
author={Ilya, L. and Frank, H.},
booktitle={Proc. of ICLR},
pages={268--295},
year={2019},
}

@inproceedings{hou2025sound,
  title={Sound-based recognition of touch gestures and emotions for enhanced human-robot interaction},
  author={Hou, Yuanbo and Ren, Qiaoqiao and Wang, Wenwu and Botteldooren, Dick},
  booktitle={IEEE International Conference on Acoustics, Speech and Signal Processing},
  pages={1--5},
  year={2025}
}

@article{sinka2021humbug,
  title={{HumBug}--{An} acoustic mosquito monitoring tool for use on budget smartphones},
  author={Sinka, M. and Zilli, D. and Li, Y. and Kiskin, I. and others},
  journal={Methods in ecology and evolution},
  volume={12},
  number={10},
  pages={1848--1859},
  year={2021}
}

@article{drbiol_icassp2026,
  title={Learning Domain-Robust Bioacoustic Representations for Mosquito Species Classification with Contrastive Learning and Distribution Alignment},
  author={Hou, Yuanbo and Liu, Zhaoyi and Shen, Xin and Roberts, Stephen},
  booktitle={IEEE International Conference on Acoustics, Speech and Signal Processing (ICASSP)},
   pages={1--5},
  year={2026},
  organization={IEEE}
}
\end{document}